\documentclass[fleqn,usenatbib]{mnras}

\usepackage{newtxtext,newtxmath}
\usepackage[T1]{fontenc}
\usepackage{ae,aecompl}
\usepackage{graphicx}
\graphicspath{{Figures/}}
\usepackage{tabularx}
\usepackage{amsmath}
\usepackage{amssymb}
\usepackage{wasysym}
\usepackage{soul}
\usepackage{times}

\newcolumntype{C}{>{\centering\arraybackslash}X}
\newcolumntype{L}{>{\raggedright\arraybackslash}X}
\newcolumntype{R}{>{\raggedleft\arraybackslash}X}

\title[AGN efficiency and Cool-cores]{Enhancing AGN efficiency and cool-core formation with anisotropic thermal conduction}

\author[D. J. Barnes et al.]{David J. Barnes$^{1}$\thanks{E-mail: djbarnes@mit.edu},
Rahul Kannan$^{2}\thanks{Einstein Fellow}$,
Mark Vogelsberger$^{1}$\thanks{Alfred P. Sloan Fellow},
Christoph Pfrommer$^{3}$,\newauthor
Ewald Puchwein$^{4,5}$,
Rainer Weinberger$^{6}$,
Volker Springel$^{6,7,8}$,
R\"{u}diger Pakmor$^{6}$,\newauthor
Dylan Nelson$^{8}$,
Federico Marinacci$^{1,2}$,
Annalisa Pillepich$^{9}$,
Paul Torrey$^{1}$,
Lars Hernquist$^{2}$
\\
$^1${Department of Physics, Kavli Institute for Astrophysics and Space Research, Massachusetts Institute of Technology, Cambridge, MA 02139, USA}\\
$^2${Harvard--Smithsonian Center for Astrophysics, 60 Garden Street, Cambridge, MA 02138}\\
$^3${Leibniz-Institut f\"{u}r Astrophysik Potsdam (AIP), An der Sternwarte 16, 14482 Potsdam, Germany}\\
$^4${Institute of Astronomy, University of Cambridge, Madingley Road, Cambridge, CB3 0HA, UK}\\
$^5${Kavli Institute for Cosmology, University of Cambridge, Madingley Road, Cambridge CB3 0HA, UK}\\
$^6${Heidelberg Institute for Theoretical Studies, Schloss-Wolfsbrunnenweg 35, D-69118 Heidelberg, Germany}\\
$^7${Zentrum f{\"u}r Astronomie der Universit{\"a}t Heidelberg, ARI, M{\"o}nchhofstr. 12-14, D-69120 Heidelberg, Germany}\\
$^8${Max-Planck-Institut f{\"u}r Astrophysik, Karl-Schwarzschild-Str. 1, 85741 Garching, Germany}\\
$^9${Max-Planck-Institut f{\"u}r Astronomie, K{\"o}nigstuhl 17, 69117 Heidelberg, Germany}\\
}

\date{Accepted XXX. Received YYY; in original form ZZZ}

\pubyear{2018}

\begin{document}
\label{firstpage}
\pagerange{\pageref{firstpage}--\pageref{lastpage}}
\maketitle

\begin{abstract}
Understanding how baryonic processes shape the intracluster medium (ICM) is of critical importance to the next generation of galaxy cluster surveys. However, most models of structure formation neglect potentially important physical processes, like anisotropic thermal conduction (ATC). In this letter, we explore the impact of ATC on the prevalence of cool-cores (CCs) using $12$ pairs of magnetohydrodynamical galaxy cluster simulations, simulated using the IllustrisTNG model with and without ATC. Although the impact of ATC varies from cluster to cluster and with CC criterion, its inclusion produces a systematic shift to larger CC fractions at $z=0$ for all CC criteria considered. Additionally, the inclusion of ATC yields a flatter CC fraction redshift evolution, easing the tension with the observed evolution. With ATC included, the energy required for the central black hole to achieve self-regulation is reduced and the gas fraction in the cluster core increases, resulting in larger CC fractions. ATC makes the ICM unstable to perturbations and the increased efficiency of AGN feedback suggests that its inclusion results in a greater level of mixing in the ICM. Therefore, ATC is potentially an important physical process in reproducing the thermal structure of the ICM.
\end{abstract}

\begin{keywords}
galaxies: clusters: general -- galaxies: clusters: intracluster medium -- conduction -- methods: numerical\vspace{-0.5cm}
\end{keywords}

\section{Introduction}
\label{sec:intro}
The thermal structure of the hot gas that fills galaxy clusters, the intracluster medium (ICM), is very sensitive to the baryonic processes that are ongoing within the cluster volume. Cooling, heat transport, turbulence, plasma physics, cosmic rays, magnetic fields and energy injection from supernovae and active galactic nuclei (AGN) all combine to shape the ICM and its observable properties. A thermal structure common to many galaxy clusters is a cool core (CC). This is a cold, dense central core whose X-ray emission implies a cooling time shorter than the age of the Universe \citep[e.g.][]{Lewis2002,Peterson2003} and is often associated with a relaxed morphology. Recent observations have shown that CCs exist at high-redshift $(z>1)$ and that their properties appear to be roughly independent of redshift \citep{McDonald2017}.

Reproducing CCs in cosmological galaxy cluster simulations remains a significant challenge. Even recent numerical work with subgrid prescriptions that include feedback from AGN struggle to reproduce the observed CC fractions from unbiased, low-redshift samples. The exact level of agreement with the observations depends on the chosen CC criteria \citep[e.g.][]{Hudson2010} and the simulated samples are small ($\sim10-30$ objects), due to their computational expense \citep[e.g.][]{Rasia2015,Hahn2017}, limiting any statistically meaningful comparison. \citet{Barnes2017c} recently explored six common CC criteria using clusters from the TNG300 volume of IllustrisTNG, which provided a sample of $370$ clusters at $z=0$. They found that the IllustrisTNG model yields reasonable agreement for some criteria compared to observed low-redshift observations, although the evolution of the CC fraction with redshift was found to be steeper than the observed trend.

However, even state-of-the-art cluster formation simulations include only the minimum physical processes required to reproduce an ICM that has reasonable global properties \citep[e.g.][]{Rasia2015,McCarthy2017,Barnes2017b,Henden2018}, i.e. cooling, star formation and feedback from supernovae and AGN. The vast majority neglect important physical processes known to impact the thermal properties of the ICM, like magnetic fields \citep[e.g.][]{Carilli2002}, cosmic rays \citep[e.g.][]{Pfrommer2013,Ruszkowski2017,Jacob2017a} and thermal conduction \citep[e.g.][]{Quataert2008,Sharma2009}. The exact impact of thermal conduction in the ICM is still unclear, the strong temperature dependence of the heat flux $(Q\propto T^{7/2})$ requires the conductivity to be fine tuned \citep{Zakamska2003} and the solutions are only locally stable on scales of the field length \citep{Soker2003}. Though thermal conduction is likely not capable of completely offsetting radiative losses, it may provide part of the heating required and it makes the ICM unstable to instabilities in the presence of external turbulence \citep{Sharma2009,BanerjeeSharma2014}. In a magnetized, weakly collisional plasma, such as the ICM, anisotropic thermal conduction (ATC) is the relevant heat transport process, with heat transport perpendicular to the magnetic field suppressed. For a single cluster, \citet{Kannan2017} demonstrated that ATC led to increased mixing and AGN feedback being more efficiently coupled to the ICM, reducing the overall energy input while increasing its ability to quench star formation.

In this Letter, we explore the impact of ATC on the formation and maintenance of CCs via cosmological zoom simulations of galaxy clusters and demonstrate that difference can be linked to efficiency with which AGN feedback couples to the ICM.

\vspace{-0.6cm}
\section{Numerical method}
\label{sec:meth}
In this work we use $12$ galaxy clusters from the \textsc{aestus} simulation suite (Kannan et al. in prep.) covering the mass range $1.0\times10^{14}\,\mathrm{M}_{\astrosun}<M_{500}<2.7\times10^{15}\,\mathrm{M}_{\astrosun}$. The suite consists of zoom simulations of galaxy clusters extracted from the Millennium XXL simulation \citep{Angulo2012}, with cosmological parameters rescaled to the  measurements of \citet{PlanckXIII2016}: $\Omega_{\mathrm{M}}=0.3089$, $\Omega_{\mathrm{\Lambda}}=0.6911$, $\Omega_{\mathrm{b}}=0.0486$, $\sigma_{8}=0.8159$, $n_{\mathrm{s}}=0.9667$ and $H=100\,h\,\mathrm{km}\,\mathrm{s}^{-1}\,\mathrm{Mpc}^{-1}$ with $h=0.6774$. The high resolution region has a dark matter and initial gas mass resolution of $5.9\times10^{7}\,\mathrm{M}_{\astrosun}$ and $1.1\times10^{7}\,\mathrm{M}_{\astrosun}$, respectively. Collisionless particles, i.e. stars and dark matter, have a softening length of $1.48\,\mathrm{kpc}$ that is comoving for $z>1$ and a fixed physical length for $z\leq1$. The gas cells employ an adaptive co-moving softening length, reaching a minimum of $0.37\,\mathrm{kpc}$.

The simulations employ the IllustrisTNG galaxy formation model \citep{Weinberger2017a,Pillepich2018}, an updated version of the Illustris model \citep{Vogelsberger2013,Torrey2014,VogelsNAT2014,Vogelsberger2014,Genel2014,Sijacki2015}, that includes radiative cooling, star formation, metal enrichment, magnetic fields and feedback from supernovae and AGN. Initial results were presented in \citet{Marinacci2017},\citet{Springel2018}, \citet{Nelson2018}, \citet{Pillepich2018b}, and \citet{Naiman2018}. The metal distribution in the ICM is well matched to observed clusters \citep{Vogelsberger2018}. A census of the CC clusters in the TNG300 volume and the redshift evolution of the CC fraction was presented in \citet{Barnes2017c}.

Each cluster in the suite is run twice, a fiducial run and a run that additionally includes ATC. The numerical implementation of ATC follows the approach introduced in \citet{Kannan2016}. The value of the conduction coefficient is set to the canonical Spitzer value \citep{Spitzer1962} parallel to the magnetic field, with a maximum diffusive value of $5\times10^{31}\,\mathrm{cm}^{2}\,\mathrm{s}^{-1}$ \citep{Ruszkowski2011,YangReynolds2016}, and zero in the direction perpendicular to the field. The conduction routine is not active for star-forming gas cells that follow an imposed equation of state \citep{Springel2003}. This conduction coefficient value is likely optimistic, as plasma effects, like mirror instabilities \citep{Komarov2016} and Whistler waves \citep{Roberg-Clark2016}, potentially lead to a substantial suppression of thermal conduction. However, we set the coefficient to the canonical Spitzer value to examine the maximum impact on the thermal structure of the ICM. \citet{Gaspari2013} have previously argued that the conduction coefficient in the ICM for isotropic conduction is negligible, however isotropic conduction is incorrect in the presence of a magnetized plasma and the value of the conduction coefficient in the ICM is still unclear.

\vspace{-0.6cm}
\section{CC fractions and redshift evolution}
\label{sec:CCfracs}
\renewcommand\arraystretch{1.2}
\renewcommand{\tabcolsep}{6pt}
\begin{table}
 \caption{Table summarizing the CC criteria used in this work.}
 \begin{tabularx}{8.5cm}{l c l}
 \hline
 Criterion & Notation & CC limit \\
 \hline
 Central electron number density & $n_{\mathrm{e}}$ & $>1.5\times10^{-2}\,\mathrm{cm}^{-3}$ \\
 Central cooling time & $t_{\mathrm{cool}}$ & $<1\,\mathrm{Gyr}$ \\
 Central entropy excess & $K_{0}$ & $<30\,\mathrm{keV}\,\mathrm{cm}^{-2}$ \\
 Concentration parameter (physical) & $C_{\mathrm{phys}}$ & $>0.155$ \\
 Concentration parameter (scaled) & $C_{\mathrm{scal}}$ & $>0.5$ \\
 Cuspiness parameter & $\alpha$ & $>0.75$ \\
 \hline
 \end{tabularx}
 \label{tab:CCprops}
\end{table}
\renewcommand\arraystretch{1.0}
\renewcommand{\tabcolsep}{6pt}

The CC definitions used in this work are summarized in Table \ref{tab:CCprops} and we refer the interested reader to \citet{Barnes2017c} for a thorough description of how these criteria are calculated.

\begin{figure*}
 \includegraphics[width=0.975\textwidth]{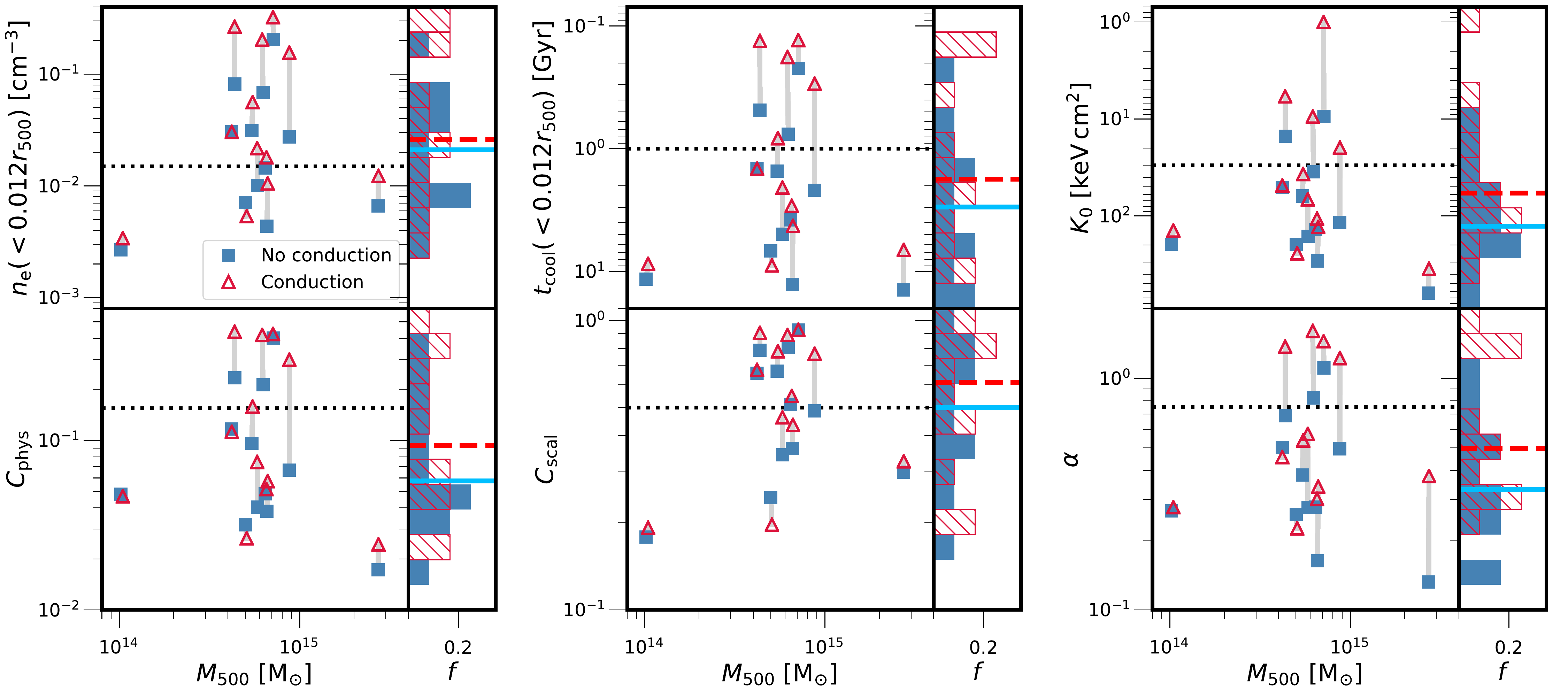}
 \vspace{-0.3cm}
 \caption{Six different CC criteria at $z=0$ for the clusters run with (red triangle) and without (blue square) anisotropic thermal conduction. Grey lines link matched clusters with the dotted black line denoting the CC threshold. We note the $y$-axis is inverted for the cooling time and central entropy excess. In the side panels the criteria distributions with (red hashed) and without (solid blue) are shown, with the median of the distributions denoted by the red dashed and blue solid lines, respectively. The inclusion of anisotropic thermal conduction produces a systematic shift towards larger CC fractions.}
\label{fig:CCs}
\end{figure*}

In the main panels of Fig. \ref{fig:CCs} we present the CC criteria as a function of $M_{500}$ at $z=0$. For a matched pair of clusters the impact of ATC varies, with several clusters showing large shifts towards the CC tail of the criterion distribution and two clusters producing small shifts towards the non-cool-core (NCC) tail. The change induced on a matched pair by ATC also varies from criterion to criterion. However, for all CC criteria explored in this work the inclusion of ATC leads to a systematic shift in the distributions towards larger CC fractions, as shown in the side panels. The fractional changes in the median CC criteria values relative to without ATC are $\Delta n_{\mathrm{e}}/n_{\mathrm{e}}=0.28$, $\Delta t_{\mathrm{cool}}/t_{\mathrm{cool}}=-0.43$, $\Delta K_{\mathrm{0}}/K_{\mathrm{0}}=-0.50$, $\Delta C_{\mathrm{phys}}/C_{\mathrm{phys}}=0.62$, $\Delta C_{\mathrm{scal}}/C_{\mathrm{scal}}=0.22$, $\Delta/\alpha=0.51$. The heat flux has a strong temperature dependence, $Q\propto T^{7/2}$, and we split the clusters by $M_{500}$ to explore this, with the $6$ most massive clusters forming a hotter sample and $6$ least massive forming a cooler sample. The inclusion of ATC produces a greater difference in the median CC criteria values of the hotter sample, with $\Delta n_{\mathrm{e}}/n_{\mathrm{e}}=0.51$, $\Delta t_{\mathrm{cool}}/t_{\mathrm{cool}}=-0.46$, $\Delta K_{\mathrm{0}}/K_{\mathrm{0}}=-0.54$, $\Delta C_{\mathrm{phys}}/C_{\mathrm{phys}}=2.10$, $\Delta C_{\mathrm{scal}}/C_{\mathrm{scal}}=0.32$, $\Delta\alpha/\alpha=1.06$, compared to the cooler sample, $\Delta n_{\mathrm{e}}/n_{\mathrm{e}}=0.24$, $\Delta t_{\mathrm{cool}}/t_{\mathrm{cool}}=-0.41$, $\Delta K_{\mathrm{0}}/K_{\mathrm{0}}=-0.48$, $\Delta C_{\mathrm{phys}}/C_{\mathrm{phys}}=0.30$, $\Delta C_{\mathrm{scal}}/C_{\mathrm{scal}}=0.14$, $\Delta\alpha/\alpha=0.41$.

\begin{figure}
 \includegraphics[width=0.95\columnwidth]{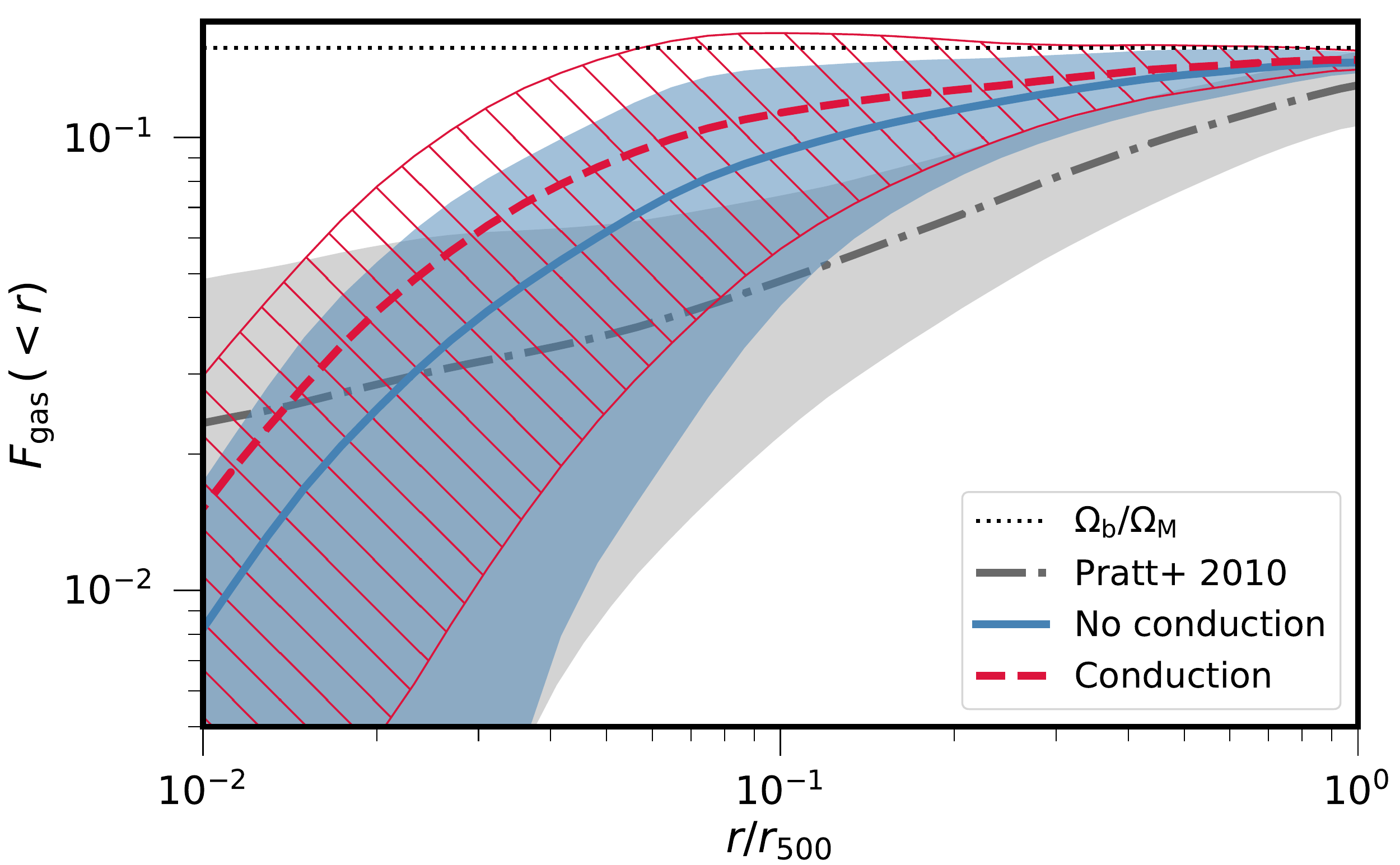}
 \vspace{-0.25cm}
 \caption{Cumulative mean gas fraction profile at $z=0$ with (red dashed) and without (blue solid) ATC. The shaded region denotes one standard deviation of the sample. The mean observed gas fraction profile (grey dash-dot) and associated standard deviation are from the REXCESS cluster sample \citep{Pratt2010}. The inclusion leads to an increased gas fraction throughout the cluster volume, but especially in the cores of clusters defined as CC.}
 \label{fig:Fgas}
\end{figure}

To further explore the shift to larger CC fractions when ATC is included, we plot the mean cumulative gas fraction radial profiles at $z=0$ for clusters with and without ATC in Fig. \ref{fig:Fgas}. We note that the shape of the gas fraction profile is inconsistent with the shape of the observed profile \citep{Pratt2010}. The inclusion of ATC leads to an increased average gas fraction throughout the cluster volume, but the difference reduces at larger radii. The fractional difference at $r_{500}$ is only a $1$ per cent increase when ATC is included, however at $0.01r_{500}$, its inclusion yields a $100$ per cent increase in the gas fraction. In addition, the inclusion of ATC leads to an increase in the scatter of the gas fraction profiles, with some profiles exceeding the universal baryon fraction $(\Omega_{\mathrm{b}}/\Omega_{\mathrm{M}})$ at $0.1r_{500}$. Defining CCs by the central electron number density criterion, all of these systems are defined as CCs. In contrast, the cumulative gas fraction profiles of NCC systems remain relatively unchanged with the inclusion of ATC.

The systematically higher gas fractions at $0.01r_{500}$ with the inclusion of ATC will result in increased central electron number densities, shifting the distribution to a greater fraction of CCs for a fixed density threshold. With $t_{\mathrm{cool}}\propto n_{\mathrm{e}}^{-1}$ and $K\propto n_{\mathrm{e}}^{-2/3}$, the increased central density will result in a shorter cooling time and a lower central entropy excess, which for a population of clusters will result in higher CC fractions. Finally, higher central densities will increase the X-ray emission from cluster cores, due to its $n_{\mathrm{e}}^{2}$ dependence, and result in larger concentration parameter values and more clusters being defined as CCs. 

\begin{figure}
 \includegraphics[width=0.95\columnwidth,height=6.45cm]{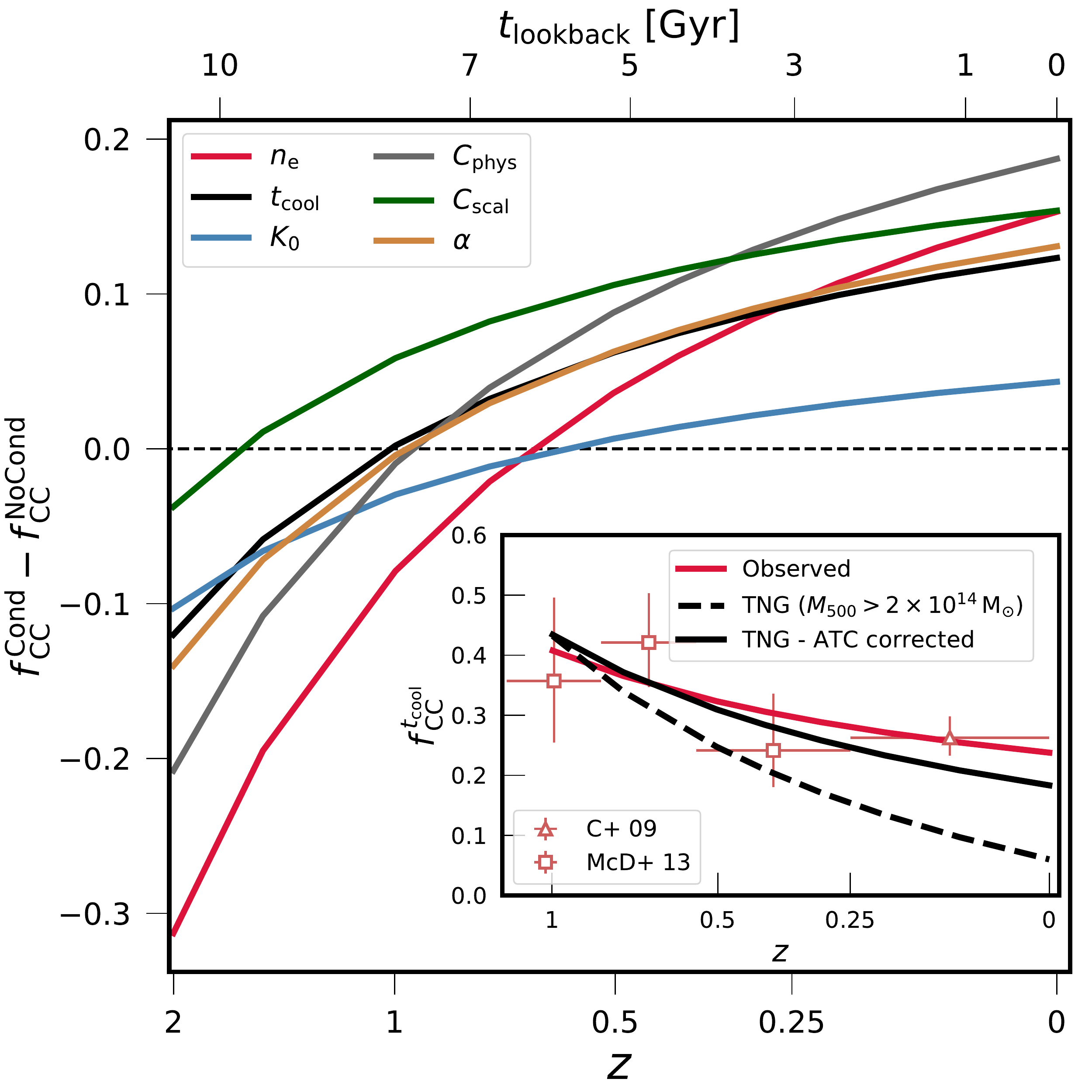}
 \vspace{-0.3cm}
 \caption{CC fraction difference with and without ATC as a function of redshift, for the central electron number density (red), cooling time (black), central entropy excess (blue), physical (grey) and scaled (green) concentration parameters and the cuspiness parameter (yellow). \textit{Inset}: Defining CCs by the central cooling time criterion, the correction factor eases the tension with the observed evolution \citep{Cavagnolo2009,McDonald2013} for the IllustrisTNG high-mass sample \citep{Barnes2017c}.}
 \label{fig:CCratio}
\end{figure}

Since the CC fraction for all 6 criteria increased at $z=0$ with the introduction of ATC, we now explore its impact on the CC fraction as a function of redshift. For all six criteria we calculate the CC fraction between $z=2$ and $z=0$, estimating the $1\sigma$ confidence intervals via the beta distribution quantile technique \citep{Cameron2011}, and then estimate the redshift evolution by fitting a linear relation that accounts for the uncertainties. We perform this fit for both samples, with and without ATC, and the difference between them is shown in the main panel of Fig. \ref{fig:CCratio}. The addition of ATC yields the same trend for all six criteria, a reduction in the CC fraction at high-redshift $(z>1)$ and an increase in the CC fraction at low-redshift. The exact change is dependent on the selected criterion, with the scaled concentration parameter showing the smallest reduction at $z=2$ $(\Delta f_{\mathrm{CC}}=-0.06)$ and one of the largest increases in CC fraction at $z=0$ $(\Delta f_{\mathrm{CC}}=0.17)$. The central entropy excess has the shallowest change between $z=2$ and $z=0$, and the smallest increase in the CC fraction at low-redshift. \citet{Barnes2017c} demonstrated that in general the IllustrisTNG model underproduced CCs at low-redshift and overproduced them at high-redshift. In the inset of Fig. \ref{fig:CCratio} we demonstrate that applying the correction factor to the redshift evolution found for such a statistically large sample, specifically the high mass $(M_{500} > 2\times10^{14}\,\mathrm{M}_{\astrosun})$ IllustrisTNG sample, eases tension with the observed redshift evolution trend for the central cooling time criterion. The observations are taken from \citet[][C+ 09]{Cavagnolo2009} and \citet[][McD+ 13]{McDonald2013}. For all criteria the addition of ATC would ease the tension with the observed redshift evolution, even if the normalization of the CC fraction still differs from observations.

\vspace{-0.68cm}
\section{The link to AGN efficiency}
\label{sec:AGNeff}
\begin{figure*}
 \includegraphics[width=0.32\textwidth]{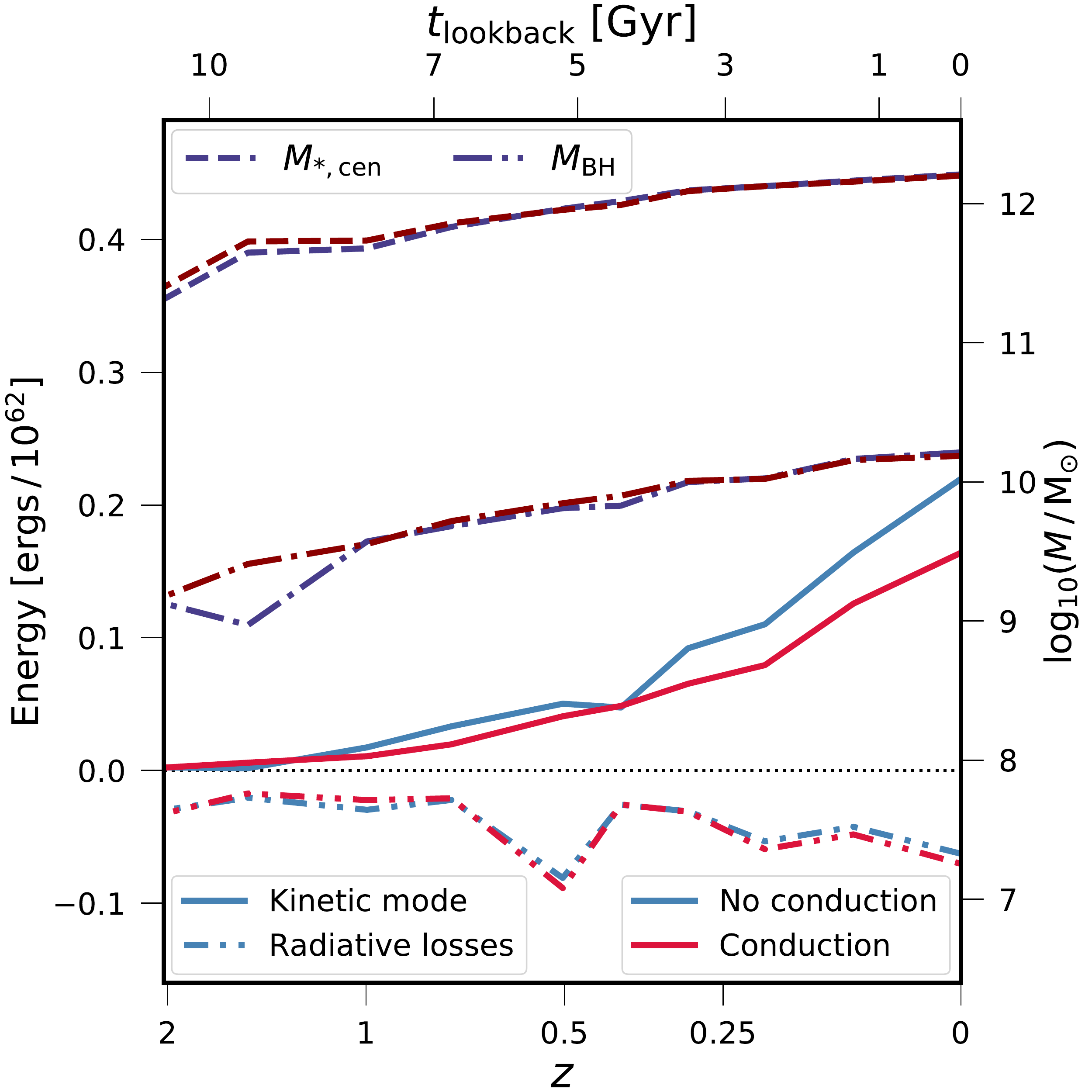}
 \includegraphics[width=0.32\textwidth]{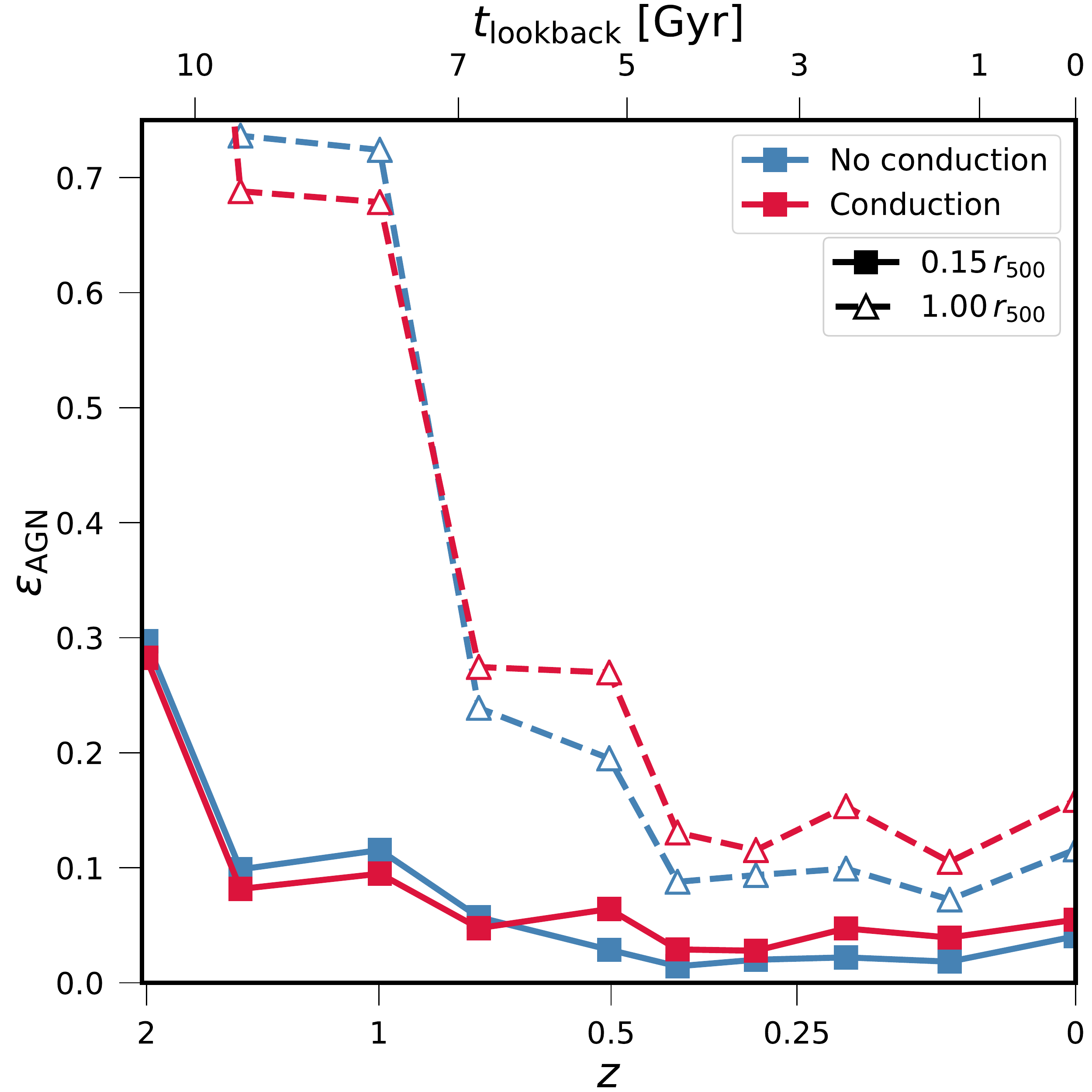}
 \includegraphics[width=0.32\textwidth]{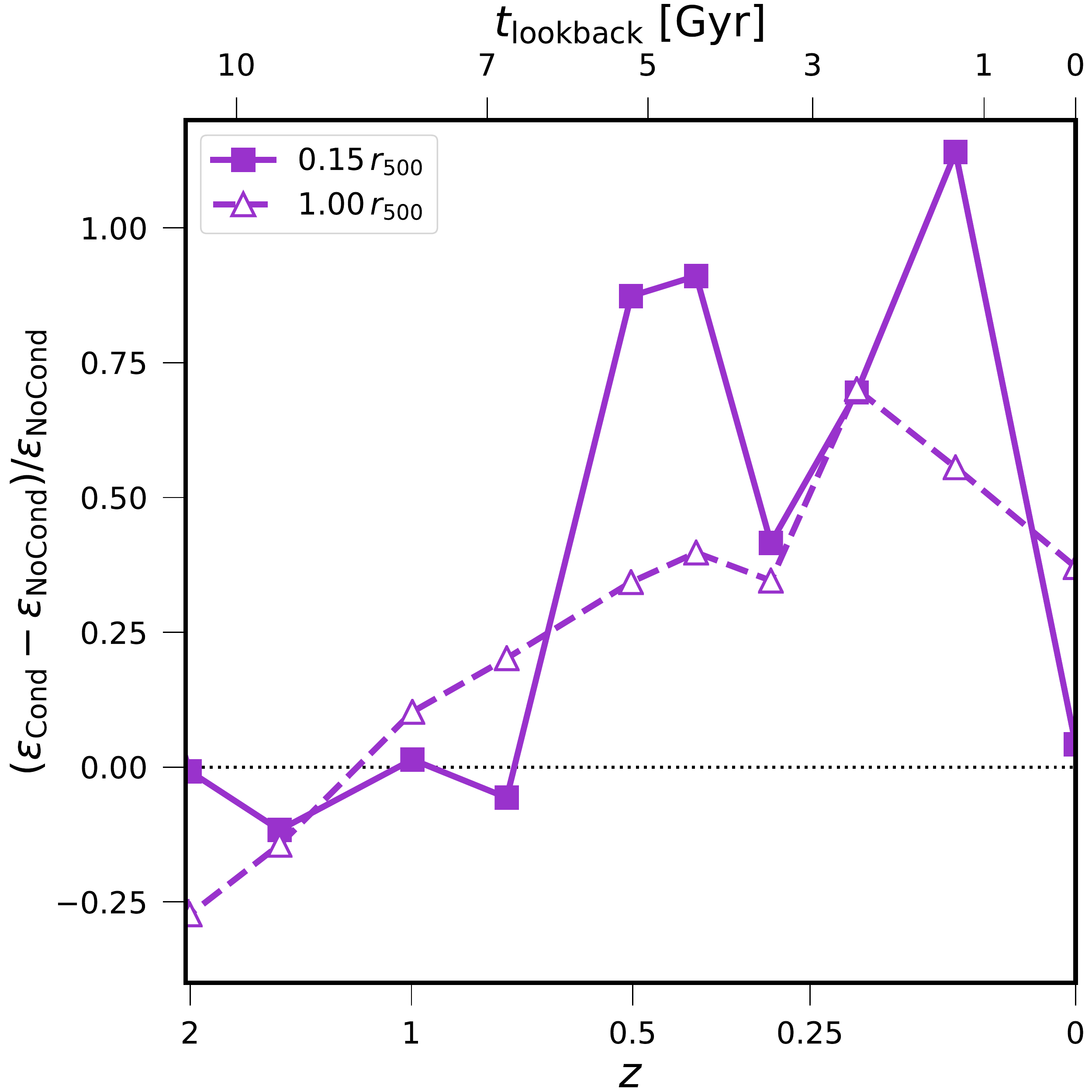}
 \vspace{-0.24cm}
 \caption{AGN feedback efficiency with (red) and without (blue) anisotropic thermal conduction. \textit{Left}: Median cumulative energy injected in the kinetic mode (solid) as a function of redshift by the black hole compared to the radiative losses within $r_{500}$ (dash-dotted). Median stellar mass of the central galaxy (dashed) and black hole mass (dash-dot) are shown by the right scale. \textit{Centre}: AGN efficiency, $\varepsilon$, measured as the ratio of radiative losses to total energy injected, as a function of redshift within apertures of $0.15r_{500}$ (solid) and $1.0r_{500}$ (dashed). \textit{Right}: Fractional difference in median AGN efficiency between simulations with and without thermal conduction. Anisotropic thermal conduction results in less energy being injected by the AGN, but a more efficient AGN-ICM coupling.}
 \label{fig:Einj}
\end{figure*}

The inclusion of ATC yields an increased CC fraction, an increased central cumulative gas fraction and a flattened redshift evolution of the CC fraction. To further understand these changes we examine the energy injected by the AGN and its coupling to the ICM. For each cluster the central black hole is defined as the most massive black hole within $1\,\mathrm{kpc}$ of the potential minimum at $z=0$ and we then traced its properties as a function of redshift. At late times the kinetic mode feedback dominates \citep{Weinberger2017c}, and the left panel of Fig. \ref{fig:Einj} shows the median energy injected by the kinetic mode as a function of redshift, with and without ATC. At $z=0$, the inclusion of ATC results in a $24$ per cent reduction in the energy injected by the low accretion rate kinetic feedback mode. The right axis shows central black hole mass and the central galaxy stellar mass, defined as the mass within twice the stellar half-mass radius of the main halo defined by \textsc{Subfind}, and the inclusion of ATC produces negligible change in both.

As an AGN will continue to inject energy until it is able to self-regulate, the lack of change in black hole or central galaxy mass and the reduction in the energy injected for simulations with ATC suggests that ATC increases the efficiency with which feedback couples to the surrounding ICM. In this work, the AGN efficiency, $\varepsilon$, is defined as the ratio of the total cumulative energy lost by the ICM via radiative cooling over the cumulative energy injected by the central AGN. When calculating the cumulative energy lost via cooling we only consider non-star-forming gas that is cooling, i.e. not being heated by supernovae or AGN feedback, and has a temperature $T\geq10^{6}\,\mathrm{K}$. Between $z=2$ and $z=0$ the cluster collapses and grows substantially and so the energy lost via cooling is measured inside spherical apertures scaled by $r_{500}$: $0.15r_{500}$ and $1.0r_{500}$. The median AGN efficiency is plotted in the central panel of Fig. \ref{fig:Einj}. Regardless of the inclusion ATC, AGN efficiency decreases towards low redshift, reducing by a factor of $\sim5$ between $z=2$ and $z=0$ for both apertures. This increased efficiency at high redshift is a combination of the AGN injecting less energy, which is seen by the steeper than linear increase in cumulative injected energy, and the increasing critical density of the Universe with redshift, which results in shorter gas cooling times (due to its $1/n$ dependence) and greater energy loss.

Finally, in the right panel of Fig. \ref{fig:Einj} we plot the fractional difference between the median AGN efficiency of the simulations with and without ATC inside apertures of $0.15r_{500}$ and $1.0r_{500}$. For $z>0.75$ the two sets of simulations have very similar AGN efficiencies within all apertures, but there is some evidence that at very high redshift $(z>1.5)$ the inclusion of ATC leads to a lower AGN efficiency. For $z<0.75$ there is a clear difference in the AGN efficiencies, with the inclusion of ATC yielding a $\sim50$ per cent increase in AGN efficiency compared to the simulations without it. This difference appears to be a function of aperture size, with the smaller aperture $(0.15r_{500})$ producing a greater difference in efficiencies compared to the larger aperture, though it is noisy.

The introduction of ATC fundamentally changes the response of the ICM to perturbations, such as AGN feedback. If a temperature gradient is present then it is unstable either to the heat-flux-driven buoyancy instability \citep{Quataert2008} or to the magnetothermal instability \citep{Balbus2000}, and any external perturbation will result in the plasma mixing \citep{Sharma2009}. \citet{Kannan2017} have already, for a single cluster, shown that the inclusion of ATC results in a greater level of mixing. The increased coupling efficiency of the AGN feedback suggests that ATC is enhancing the mixing in the ICM, distributing the kinetic mode feedback more isotropically. This requires less energy per unit time to be injected by the central black hole to offset radiative losses, which softens the AGN feedback and increases the central gas density and the measured CC fraction.

\vspace{-0.7cm}
\section{Conclusions}
\label{sec:concs}
We have examined the impact of ATC on CC formation and AGN feedback efficiency using $12$ zoom cosmological MHD galaxy cluster simulations. Each cluster was run with and without ATC using the IllustrisTNG model. ATC was self-consistently included following the method of \citet{Kannan2016}. Our main results are:
\begin{itemize}
 \item Although the impact of ATC varied on a cluster by cluster and on a CC criterion basis, its inclusion produced a systematic shift of all criteria distributions to larger CC fractions (Fig. \ref{fig:CCs}). This is the result of an increased central gas fraction with ATC (Fig. \ref{fig:Fgas}).
 \item Exploring the redshift evolution of the CC fraction, we found that inclusion of ATC resulted in a decreased CC fraction at high redshift ($z\geq1$) and an increased CC fraction at low redshift ($z<1$) (Fig. \ref{fig:CCratio}). For the statistically large sample produced by IllustrisTNG \citep{Barnes2017c}, the correction due to ATC would ease the tension between the simulated and observed redshift evolution of the CC fraction.
 \item Finally, we examined the total energy injected by the central black hole and found that inclusion of ATC resulted in a $7$ per cent decrease in the total energy injected, due to the kinetic feedback mode injecting $24$ per cent less energy during the formation of the cluster (Fig. \ref{fig:Einj}). The increased efficiency of AGN feedback suggests that ATC increases the mixing in the ICM, softening the impact of AGN feedback and producing higher central gas fractions.
\end{itemize}
The results suggest that the inclusion of ATC alters the response of the ICM to AGN feedback, making it unstable to convective mixing in the presence of external perturbations. The increased mixing makes the AGN-ICM coupling more efficient, reducing the energy required for self-regulation \citep[e.g.][]{Kannan2017}. This results in increased low-redshift CC fractions and flatter redshift evolution compared to the same model without ATC. Previous numerical work with artificial thermal conduction that promoted mixing also found an improved match to observed CC fractions \citep{Rasia2015}. The inclusion of ATC reduces the entropy injected by a self-regulating black hole, suggesting its inclusion may help numerical simulations reproduce the thermal structure of cluster cores.

\vspace{-0.65cm}
\section*{Acknowledgements}
The simulations were performed on the MKI-Harvard Odyssey cluster and the Stampede supercomputers at the Texas Advanced Computing Center as part of XSEDE project TG-AST160069. RK acknowledges support from NASA  through  Einstein  Postdoctoral  Fellowship  grant  number PF7-180163 awarded  by  the  \textit{Chandra} X-ray  Center,  which  is  operated  by  the  Smithsonian Astrophysical Observatory for NASA under contract NAS8-03060. MV acknowledges support through an MIT RSC award, the support of the Alfred P. Sloan Foundation, and support by NASA ATP grant NNX17AG29G. PT acknowledges support from NASA through Hubble Fellowship grants HST-HF2-51384.001-A awarded by the STScI, which is operated by the Association of Universities for Research in Astronomy, Inc., for NASA, under contract NAS5-26555. CP acknowledges support by the European Research Council under ERC-CoG grant CRAGSMAN-646955.
\vspace{-0.6cm}

%%%%%%%%%%%%%%%%%%%% REFERENCES %%%%%%%%%%%%%%%%%%
\bibliographystyle{mnras}
\bibliography{main}

\newcommand{\noop}[1]{}
\begin{thebibliography}{}
\makeatletter
\relax
\def\mn@urlcharsother{\let\do\@makeother \do\$\do\&\do\#\do\^\do\_\do\%\do\~}
\def\mn@doi{\begingroup\mn@urlcharsother \@ifnextchar [ {\mn@doi@}
  {\mn@doi@[]}}
\def\mn@doi@[#1]#2{\def\@tempa{#1}\ifx\@tempa\@empty \href
  {http://dx.doi.org/#2} {doi:#2}\else \href {http://dx.doi.org/#2} {#1}\fi
  \endgroup}
\def\mn@eprint#1#2{\mn@eprint@#1:#2::\@nil}
\def\mn@eprint@arXiv#1{\href {http://arxiv.org/abs/#1} {{\tt arXiv:#1}}}
\def\mn@eprint@dblp#1{\href {http://dblp.uni-trier.de/rec/bibtex/#1.xml}
  {dblp:#1}}
\def\mn@eprint@#1:#2:#3:#4\@nil{\def\@tempa {#1}\def\@tempb {#2}\def\@tempc
  {#3}\ifx \@tempc \@empty \let \@tempc \@tempb \let \@tempb \@tempa \fi \ifx
  \@tempb \@empty \def\@tempb {arXiv}\fi \@ifundefined
  {mn@eprint@\@tempb}{\@tempb:\@tempc}{\expandafter \expandafter \csname
  mn@eprint@\@tempb\endcsname \expandafter{\@tempc}}}

\bibitem[\protect\citeauthoryear{{Angulo}, {Springel}, {White}, {Jenkins},
  {Baugh}  \& {Frenk}}{{Angulo} et~al.}{2012}]{Angulo2012}
{Angulo} R.~E.,  {Springel} V.,  {White} S.~D.~M.,  {Jenkins} A.,  {Baugh}
  C.~M.,   {Frenk} C.~S.,  2012, \mn@doi [\mnras]
  {10.1111/j.1365-2966.2012.21830.x}, \href
  {http://adsabs.harvard.edu/abs/2012MNRAS.426.2046A} {426, 2046}

\bibitem[\protect\citeauthoryear{{Balbus}}{{Balbus}}{2000}]{Balbus2000}
{Balbus} S.~A.,  2000, \mn@doi [\apj] {10.1086/308732}, \href
  {http://adsabs.harvard.edu/abs/2000ApJ...534..420B} {534, 420}

\bibitem[\protect\citeauthoryear{{Banerjee} \& {Sharma}}{{Banerjee} \&
  {Sharma}}{2014}]{BanerjeeSharma2014}
{Banerjee} N.,  {Sharma} P.,  2014, \mn@doi [\mnras] {10.1093/mnras/stu1179},
  \href {http://adsabs.harvard.edu/abs/2014MNRAS.443..687B} {443, 687}

\bibitem[\protect\citeauthoryear{{Barnes} et~al.,}{{Barnes}
  et~al.}{2017a}]{Barnes2017c}
{Barnes} D.~J.,  et~al., 2017a, preprint, \href
  {http://adsabs.harvard.edu/abs/2017arXiv171008420B} {} (\mn@eprint {arXiv}
  {1710.08420})

\bibitem[\protect\citeauthoryear{{Barnes} et~al.,}{{Barnes}
  et~al.}{2017b}]{Barnes2017b}
{Barnes} D.~J.,  et~al., 2017b, \mn@doi [\mnras] {10.1093/mnras/stx1647}, \href
  {http://adsabs.harvard.edu/abs/2017MNRAS.471.1088B} {471, 1088}

\bibitem[\protect\citeauthoryear{{Cameron}}{{Cameron}}{2011}]{Cameron2011}
{Cameron} E.,  2011, \mn@doi [\pasa] {10.1071/AS10046}, \href
  {http://adsabs.harvard.edu/abs/2011PASA...28..128C} {28, 128}

\bibitem[\protect\citeauthoryear{{Carilli} \& {Taylor}}{{Carilli} \&
  {Taylor}}{2002}]{Carilli2002}
{Carilli} C.~L.,  {Taylor} G.~B.,  2002, \mn@doi [\araa]
  {10.1146/annurev.astro.40.060401.093852}, \href
  {http://adsabs.harvard.edu/abs/2002ARA%26A..40..319C} {40, 319}

\bibitem[\protect\citeauthoryear{{Cavagnolo}, {Donahue}, {Voit}  \&
  {Sun}}{{Cavagnolo} et~al.}{2009}]{Cavagnolo2009}
{Cavagnolo} K.~W.,  {Donahue} M.,  {Voit} G.~M.,   {Sun} M.,  2009, \mn@doi
  [\apjs] {10.1088/0067-0049/182/1/12}, \href
  {http://adsabs.harvard.edu/abs/2009ApJS..182...12C} {182, 12}

\bibitem[\protect\citeauthoryear{{Gaspari} \& {Churazov}}{{Gaspari} \&
  {Churazov}}{2013}]{Gaspari2013}
{Gaspari} M.,  {Churazov} E.,  2013, \mn@doi [\aap]
  {10.1051/0004-6361/201322295}, \href
  {http://adsabs.harvard.edu/abs/2013A%26A...559A..78G} {559, A78}

\bibitem[\protect\citeauthoryear{{Genel} et~al.,}{{Genel}
  et~al.}{2014}]{Genel2014}
{Genel} S.,  et~al., 2014, \mn@doi [\mnras] {10.1093/mnras/stu1654}, \href
  {http://adsabs.harvard.edu/abs/2014MNRAS.445..175G} {445, 175}

\bibitem[\protect\citeauthoryear{{Hahn}, {Martizzi}, {Wu}, {Evrard}, {Teyssier}
   \& {Wechsler}}{{Hahn} et~al.}{2017}]{Hahn2017}
{Hahn} O.,  {Martizzi} D.,  {Wu} H.-Y.,  {Evrard} A.~E.,  {Teyssier} R.,
  {Wechsler} R.~H.,  2017, \mn@doi [\mnras] {10.1093/mnras/stx001}, \href
  {http://adsabs.harvard.edu/abs/2017MNRAS.470..166H} {470, 166}

\bibitem[\protect\citeauthoryear{{Henden}, {Puchwein}, {Shen}  \&
  {Sijacki}}{{Henden} et~al.}{2018}]{Henden2018}
{Henden} N.~A.,  {Puchwein} E.,  {Shen} S.,   {Sijacki} D.,  2018, preprint,
  \href {http://adsabs.harvard.edu/abs/2018arXiv180405064H} {} (\mn@eprint
  {arXiv} {1804.05064})

\bibitem[\protect\citeauthoryear{{Hudson}, {Mittal}, {Reiprich}, {Nulsen},
  {Andernach}  \& {Sarazin}}{{Hudson} et~al.}{2010}]{Hudson2010}
{Hudson} D.~S.,  {Mittal} R.,  {Reiprich} T.~H.,  {Nulsen} P.~E.~J.,
  {Andernach} H.,   {Sarazin} C.~L.,  2010, \mn@doi [\aap]
  {10.1051/0004-6361/200912377}, \href
  {http://adsabs.harvard.edu/abs/2010A%26A...513A..37H} {513, A37}

\bibitem[\protect\citeauthoryear{{Jacob} \& {Pfrommer}}{{Jacob} \&
  {Pfrommer}}{2017}]{Jacob2017a}
{Jacob} S.,  {Pfrommer} C.,  2017, \mn@doi [\mnras] {10.1093/mnras/stx131},
  \href {http://adsabs.harvard.edu/abs/2017MNRAS.467.1449J} {467, 1449}

\bibitem[\protect\citeauthoryear{{Kannan}, {Springel}, {Pakmor}, {Marinacci}
  \& {Vogelsberger}}{{Kannan} et~al.}{2016}]{Kannan2016}
{Kannan} R.,  {Springel} V.,  {Pakmor} R.,  {Marinacci} F.,   {Vogelsberger}
  M.,  2016, \mn@doi [\mnras] {10.1093/mnras/stw294}, \href
  {http://adsabs.harvard.edu/abs/2016MNRAS.458..410K} {458, 410}

\bibitem[\protect\citeauthoryear{{Kannan}, {Vogelsberger}, {Pfrommer},
  {Weinberger}, {Springel}, {Hernquist}, {Puchwein}  \& {Pakmor}}{{Kannan}
  et~al.}{2017}]{Kannan2017}
{Kannan} R.,  {Vogelsberger} M.,  {Pfrommer} C.,  {Weinberger} R.,  {Springel}
  V.,  {Hernquist} L.,  {Puchwein} E.,   {Pakmor} R.,  2017, \mn@doi [\apjl]
  {10.3847/2041-8213/aa624b}, \href
  {http://adsabs.harvard.edu/abs/2017ApJ...837L..18K} {837, L18}

\bibitem[\protect\citeauthoryear{{Komarov}, {Churazov}, {Kunz}  \&
  {Schekochihin}}{{Komarov} et~al.}{2016}]{Komarov2016}
{Komarov} S.~V.,  {Churazov} E.~M.,  {Kunz} M.~W.,   {Schekochihin} A.~A.,
  2016, \mn@doi [\mnras] {10.1093/mnras/stw963}, \href
  {http://adsabs.harvard.edu/abs/2016MNRAS.460..467K} {460, 467}

\bibitem[\protect\citeauthoryear{{Lewis}, {Stocke}  \& {Buote}}{{Lewis}
  et~al.}{2002}]{Lewis2002}
{Lewis} A.~D.,  {Stocke} J.~T.,   {Buote} D.~A.,  2002, \mn@doi [\apjl]
  {10.1086/341990}, \href {http://adsabs.harvard.edu/abs/2002ApJ...573L..13L}
  {573, L13}

\bibitem[\protect\citeauthoryear{{Marinacci} et~al.,}{{Marinacci}
  et~al.}{2017}]{Marinacci2017}
{Marinacci} F.,  et~al., 2017, preprint, \href
  {http://adsabs.harvard.edu/abs/2017arXiv170703396M} {} (\mn@eprint {arXiv}
  {1707.03396})

\bibitem[\protect\citeauthoryear{{McCarthy}, {Schaye}, {Bird}  \& {Le
  Brun}}{{McCarthy} et~al.}{2017}]{McCarthy2017}
{McCarthy} I.~G.,  {Schaye} J.,  {Bird} S.,   {Le Brun} A.~M.~C.,  2017,
  \mn@doi [\mnras] {10.1093/mnras/stw2792}, \href
  {http://adsabs.harvard.edu/abs/2017MNRAS.465.2936M} {465, 2936}

\bibitem[\protect\citeauthoryear{{McDonald} et~al.,}{{McDonald}
  et~al.}{2013}]{McDonald2013}
{McDonald} M.,  et~al., 2013, \mn@doi [\apj] {10.1088/0004-637X/774/1/23},
  \href {http://adsabs.harvard.edu/abs/2013ApJ...774...23M} {774, 23}

\bibitem[\protect\citeauthoryear{{McDonald} et~al.,}{{McDonald}
  et~al.}{2017}]{McDonald2017}
{McDonald} M.,  et~al., 2017, \mn@doi [\apj] {10.3847/1538-4357/aa7740}, \href
  {http://adsabs.harvard.edu/abs/2017ApJ...843...28M} {843, 28}

\bibitem[\protect\citeauthoryear{{Naiman} et~al.,}{{Naiman}
  et~al.}{2018}]{Naiman2018}
{Naiman} J.~P.,  et~al., 2018, \mn@doi [\mnras] {10.1093/mnras/sty618}, \href
  {http://adsabs.harvard.edu/abs/2018MNRAS.tmp..585N} {}

\bibitem[\protect\citeauthoryear{{Nelson} et~al.,}{{Nelson}
  et~al.}{2018}]{Nelson2018}
{Nelson} D.,  et~al., 2018, \mn@doi [\mnras] {10.1093/mnras/stx3040}, \href
  {http://adsabs.harvard.edu/abs/2018MNRAS.475..624N} {475, 624}

\bibitem[\protect\citeauthoryear{{Peterson}, {Kahn}, {Paerels}, {Kaastra},
  {Tamura}, {Bleeker}, {Ferrigno}  \& {Jernigan}}{{Peterson}
  et~al.}{2003}]{Peterson2003}
{Peterson} J.~R.,  {Kahn} S.~M.,  {Paerels} F.~B.~S.,  {Kaastra} J.~S.,
  {Tamura} T.,  {Bleeker} J.~A.~M.,  {Ferrigno} C.,   {Jernigan} J.~G.,  2003,
  \mn@doi [\apj] {10.1086/374830}, \href
  {http://adsabs.harvard.edu/abs/2003ApJ...590..207P} {590, 207}

\bibitem[\protect\citeauthoryear{{Pfrommer}}{{Pfrommer}}{2013}]{Pfrommer2013}
{Pfrommer} C.,  2013, \mn@doi [\apj] {10.1088/0004-637X/779/1/10}, \href
  {http://adsabs.harvard.edu/abs/2013ApJ...779...10P} {779, 10}

\bibitem[\protect\citeauthoryear{{Pillepich} et~al.,}{{Pillepich}
  et~al.}{2018a}]{Pillepich2018}
{Pillepich} A.,  et~al., 2018a, \mn@doi [\mnras] {10.1093/mnras/stx2656}, \href
  {http://adsabs.harvard.edu/abs/2018MNRAS.473.4077P} {473, 4077}

\bibitem[\protect\citeauthoryear{{Pillepich} et~al.,}{{Pillepich}
  et~al.}{2018b}]{Pillepich2018b}
{Pillepich} A.,  et~al., 2018b, \mn@doi [\mnras] {10.1093/mnras/stx3112}, \href
  {http://adsabs.harvard.edu/abs/2018MNRAS.475..648P} {475, 648}

\bibitem[\protect\citeauthoryear{{Planck Collaboration} et~al.,}{{Planck
  Collaboration} et~al.}{2016}]{PlanckXIII2016}
{Planck Collaboration} et~al., 2016, \mn@doi [\aap]
  {10.1051/0004-6361/201525830}, \href
  {http://adsabs.harvard.edu/abs/2016A%26A...594A..13P} {594, A13}

\bibitem[\protect\citeauthoryear{{Pratt} et~al.,}{{Pratt}
  et~al.}{2010}]{Pratt2010}
{Pratt} G.~W.,  et~al., 2010, \mn@doi [\aap] {10.1051/0004-6361/200913309},
  \href {http://adsabs.harvard.edu/abs/2010A%26A...511A..85P} {511, A85}

\bibitem[\protect\citeauthoryear{{Quataert}}{{Quataert}}{2008}]{Quataert2008}
{Quataert} E.,  2008, \mn@doi [\apj] {10.1086/525248}, \href
  {http://adsabs.harvard.edu/abs/2008ApJ...673..758Q} {673, 758}

\bibitem[\protect\citeauthoryear{{Rasia} et~al.,}{{Rasia}
  et~al.}{2015}]{Rasia2015}
{Rasia} E.,  et~al., 2015, \mn@doi [\apjl] {10.1088/2041-8205/813/1/L17}, \href
  {http://adsabs.harvard.edu/abs/2015ApJ...813L..17R} {813, L17}

\bibitem[\protect\citeauthoryear{{Roberg-Clark}, {Drake}, {Reynolds}  \&
  {Swisdak}}{{Roberg-Clark} et~al.}{2016}]{Roberg-Clark2016}
{Roberg-Clark} G.~T.,  {Drake} J.~F.,  {Reynolds} C.~S.,   {Swisdak} M.,  2016,
  \mn@doi [\apjl] {10.3847/2041-8205/830/1/L9}, \href
  {http://adsabs.harvard.edu/abs/2016ApJ...830L...9R} {830, L9}

\bibitem[\protect\citeauthoryear{{Ruszkowski}, {Lee}, {Br{\"u}ggen}, {Parrish}
  \& {Oh}}{{Ruszkowski} et~al.}{2011}]{Ruszkowski2011}
{Ruszkowski} M.,  {Lee} D.,  {Br{\"u}ggen} M.,  {Parrish} I.,   {Oh} S.~P.,
  2011, \mn@doi [\apj] {10.1088/0004-637X/740/2/81}, \href
  {http://adsabs.harvard.edu/abs/2011ApJ...740...81R} {740, 81}

\bibitem[\protect\citeauthoryear{{Ruszkowski}, {Yang}  \&
  {Reynolds}}{{Ruszkowski} et~al.}{2017}]{Ruszkowski2017}
{Ruszkowski} M.,  {Yang} H.-Y.~K.,   {Reynolds} C.~S.,  2017, \mn@doi [\apj]
  {10.3847/1538-4357/aa79f8}, \href
  {http://adsabs.harvard.edu/abs/2017ApJ...844...13R} {844, 13}

\bibitem[\protect\citeauthoryear{{Sharma}, {Chandran}, {Quataert}  \&
  {Parrish}}{{Sharma} et~al.}{2009}]{Sharma2009}
{Sharma} P.,  {Chandran} B.~D.~G.,  {Quataert} E.,   {Parrish} I.~J.,  2009,
  \mn@doi [\apj] {10.1088/0004-637X/699/1/348}, \href
  {http://adsabs.harvard.edu/abs/2009ApJ...699..348S} {699, 348}

\bibitem[\protect\citeauthoryear{{Sijacki}, {Vogelsberger}, {Genel},
  {Springel}, {Torrey}, {Snyder}, {Nelson}  \& {Hernquist}}{{Sijacki}
  et~al.}{2015}]{Sijacki2015}
{Sijacki} D.,  {Vogelsberger} M.,  {Genel} S.,  {Springel} V.,  {Torrey} P.,
  {Snyder} G.~F.,  {Nelson} D.,   {Hernquist} L.,  2015, \mn@doi [\mnras]
  {10.1093/mnras/stv1340}, \href
  {http://adsabs.harvard.edu/abs/2015MNRAS.452..575S} {452, 575}

\bibitem[\protect\citeauthoryear{{Soker}}{{Soker}}{2003}]{Soker2003}
{Soker} N.,  2003, \mn@doi [\mnras] {10.1046/j.1365-8711.2003.06548.x}, \href
  {http://adsabs.harvard.edu/abs/2003MNRAS.342..463S} {342, 463}

\bibitem[\protect\citeauthoryear{{Spitzer}}{{Spitzer}}{1962}]{Spitzer1962}
{Spitzer} L.,  1962, {Physics of Fully Ionized Gases}

\bibitem[\protect\citeauthoryear{{Springel} \& {Hernquist}}{{Springel} \&
  {Hernquist}}{2003}]{Springel2003}
{Springel} V.,  {Hernquist} L.,  2003, \mn@doi [\mnras]
  {10.1046/j.1365-8711.2003.06206.x}, \href
  {http://adsabs.harvard.edu/abs/2003MNRAS.339..289S} {339, 289}

\bibitem[\protect\citeauthoryear{{Springel} et~al.,}{{Springel}
  et~al.}{2018}]{Springel2018}
{Springel} V.,  et~al., 2018, \mn@doi [\mnras] {10.1093/mnras/stx3304}, \href
  {http://adsabs.harvard.edu/abs/2018MNRAS.475..676S} {475, 676}

\bibitem[\protect\citeauthoryear{{Torrey}, {Vogelsberger}, {Genel}, {Sijacki},
  {Springel}  \& {Hernquist}}{{Torrey} et~al.}{2014}]{Torrey2014}
{Torrey} P.,  {Vogelsberger} M.,  {Genel} S.,  {Sijacki} D.,  {Springel} V.,
  {Hernquist} L.,  2014, \mn@doi [\mnras] {10.1093/mnras/stt2295}, \href
  {http://adsabs.harvard.edu/abs/2014MNRAS.438.1985T} {438, 1985}

\bibitem[\protect\citeauthoryear{{Vogelsberger}, {Genel}, {Sijacki}, {Torrey},
  {Springel}  \& {Hernquist}}{{Vogelsberger} et~al.}{2013}]{Vogelsberger2013}
{Vogelsberger} M.,  {Genel} S.,  {Sijacki} D.,  {Torrey} P.,  {Springel} V.,
  {Hernquist} L.,  2013, \mn@doi [\mnras] {10.1093/mnras/stt1789}, \href
  {http://adsabs.harvard.edu/abs/2013MNRAS.436.3031V} {436, 3031}

\bibitem[\protect\citeauthoryear{{Vogelsberger} et~al.,}{{Vogelsberger}
  et~al.}{2014a}]{Vogelsberger2014}
{Vogelsberger} M.,  et~al., 2014a, \mn@doi [\mnras] {10.1093/mnras/stu1536},
  \href {http://adsabs.harvard.edu/abs/2014MNRAS.444.1518V} {444, 1518}

\bibitem[\protect\citeauthoryear{{Vogelsberger} et~al.,}{{Vogelsberger}
  et~al.}{2014b}]{VogelsNAT2014}
{Vogelsberger} M.,  et~al., 2014b, \mn@doi [\nat] {10.1038/nature13316}, \href
  {http://adsabs.harvard.edu/abs/2014Natur.509..177V} {509, 177}

\bibitem[\protect\citeauthoryear{{Vogelsberger} et~al.,}{{Vogelsberger}
  et~al.}{2018}]{Vogelsberger2018}
{Vogelsberger} M.,  et~al., 2018, \mn@doi [\mnras] {10.1093/mnras/stx2955},
  \href {http://adsabs.harvard.edu/abs/2018MNRAS.474.2073V} {474, 2073}

\bibitem[\protect\citeauthoryear{{Weinberger} et~al.,}{{Weinberger}
  et~al.}{2017a}]{Weinberger2017c}
{Weinberger} R.,  et~al., 2017a, preprint, \href
  {http://adsabs.harvard.edu/abs/2017arXiv171004659W} {} (\mn@eprint {arXiv}
  {1710.04659})

\bibitem[\protect\citeauthoryear{{Weinberger} et~al.,}{{Weinberger}
  et~al.}{2017b}]{Weinberger2017a}
{Weinberger} R.,  et~al., 2017b, \mn@doi [\mnras] {10.1093/mnras/stw2944},
  \href {http://adsabs.harvard.edu/abs/2017MNRAS.465.3291W} {465, 3291}

\bibitem[\protect\citeauthoryear{{Yang} \& {Reynolds}}{{Yang} \&
  {Reynolds}}{2016}]{YangReynolds2016}
{Yang} H.-Y.~K.,  {Reynolds} C.~S.,  2016, \mn@doi [\apj]
  {10.3847/0004-637X/829/2/90}, \href
  {http://adsabs.harvard.edu/abs/2016ApJ...829...90Y} {829, 90}

\bibitem[\protect\citeauthoryear{{Zakamska} \& {Narayan}}{{Zakamska} \&
  {Narayan}}{2003}]{Zakamska2003}
{Zakamska} N.~L.,  {Narayan} R.,  2003, \mn@doi [\apj] {10.1086/344641}, \href
  {http://adsabs.harvard.edu/abs/2003ApJ...582..162Z} {582, 162}

\makeatother
\end{thebibliography}

\bsp	
\label{lastpage}
\end{document}